\documentclass[twocolumn,prl,amsmath,amssymb,showpacs,
superscriptaddress,floatfix]{revtex4}

\usepackage{graphicx}

\begin{document}

\title{Andreev conductance of a domain wall}
\author{Nikolai~M. Chtchelkatchev}

%\email{nms@landau.ac.ru}
\affiliation{L.D.\ Landau Institute for Theoretical Physics,
Russian Academy of Sciences, 117940 Moscow, Russia}
\affiliation{Institute for High Pressure Physics, Russian Academy
of Sciences, Troitsk 142092, Moscow Region, Russia}

\author{Igor~S. Burmistrov}
%\email{Y.V.Nazarov@tn.tudelft.nl}
\affiliation{L.D.\ Landau Institute for Theoretical Physics,
Russian Academy of Sciences, 117940 Moscow, Russia}
%\affiliation{Institute for Theoretical Physics, University of
%Amsterdam, Valckenierstraat 65, 1018 XE Amsterdam, The
%Netherlands}

\date{\today}

\begin{abstract}
At low temperatures, the transport through a
superconductor-ferromagnet tunnel interface is due to tunneling of
electrons in pairs. Exchange field of a monodomain ferromagnet
aligns electron spins and suppresses the two electron tunneling.
The presence of the domain walls at the SF interface strongly
enhances the subgap current. The Andreev conductance is proven to
be proportional to the total length of domain walls at the SF
interface.
\end{abstract}


\maketitle

Interplay of superconductivity and ferromagnetism in mesoscopic
hybrid structures is now in focus of both experimental and
theoretical research. Superconductivity and ferromagnetism are two
competing orders: while the former prefers antiparallel spin
orientation of electrons in Cooper pairs, the latter forces the
spins to align in parallel. Their coexistence in one and the same
material or when the two interactions are spatially separated
leads to a number of new interesting effects such as $\pi$-state
of SFS Josephson junctions \cite{Bulaevskii,Ryazanov,Chtch_pi},
highly nonmonotonic dependence of the critical temperature of the
SF system on the thickness of ferromagnet
\cite{Fominov_Ch_Golubov}, \textit{etc}. Investigations of the SF
structures are often based on a bare assumption that ferromagnet
consists of the only domain or that a domain structure is not
important. However, this approximation is not always valid
\cite{Sonin,Pokrovsky,Kinsey,Ryazanov_vortices}. Recently it was
demonstrated that due to a domain structure of the ferromagnet in
SF bilayer vortices may appear in the superconducting film  and
significantly modify lateral conductance of this system
\cite{Ryazanov_vortices}.

This paper is largely concerned with the influence of the
ferromagnetic domain structure on the Andreev conductance of SF
junctions. First consider the superconductor-ferromagnet junction
with a monodomain ferromagnet. When the voltage $V$ between the
superconductor and the ferromagnet is smaller than the
superconducting gap $\Delta$ an electron exchange between
superconductor and ferromagnet is provided by the Andreev
processes \cite{Andreev}. They involve transfer of two electrons
with opposite spin from the ferromagnet into the superconductor or
vice-versa. The Andreev conductance is proportional therefore to
the product $ \nu_{\nparallel}\nu_\parallel$ of the minority
$\nu_{\nparallel}$ and majority $\nu_{\parallel}$ band densities
of states in the ferromagnet. Thus when in the ferromagnet  the
majority of electron spins are polarized along the direction of
the magnetization subgap electron transport through the SF
junction is suppressed.

\begin{figure}[t]
\includegraphics[width=50mm]{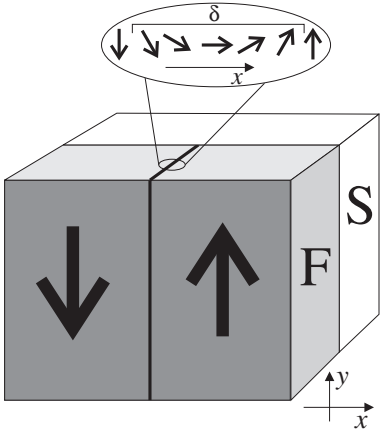}
\caption{\label{fig1} The superconductor ferromagnet junction. The
domain wall.
%The normal conductance between the superconductor and the
%ferromagnet in an experimental setup should be smaller than the
%normal conductances of the leads $L_{1,2}$ and the ferromagnet
%(superconductor); this condition allows to believe that F and S
%are in equilibrium and measure the SF conductance rather than
%lead-related effects.
}
\end{figure}

If the ferromagnet consists of several domains, domain walls
separate regions with different direction of magnetization. When a
domain wall is situated near the SF interface electrons with
opposite spins involved in the Andreev processes origin from the
adjacent domains. This effect makes the Andreev conductance finite
at any polarization of the ferromagnet. In the case of the fully
polarized ferromagnet ($\nu_{\nparallel}\ll \nu_\parallel$) we
derived that the Andreev conductance of the SF junction is
proportional to the total length $L_D^{(\rm tot)}$ of the domain
walls situated at the SF boundary and is given by
\begin{gather} \label{1} G_{A}=
\frac{\hbar}{\pi e^{2}}\frac{g_{_N}^2}{\nu_S\Delta}L_D^{(\rm tot)}
f\left (\frac{\pi^{2} \xi_{0}}{16 \delta} \right )\, ,
\end{gather}
where $\nu_S$ is the density of states in the superconductor,
$g_{_N}$ stands for the normal conductance of the SF junction per
unit area, and $\delta$ is the width of the domain wall [see
Fig.~\ref{fig1}]. The coherence length of the superconductor
$\xi_{0}$ is equaled to $v_{F}/\pi \Delta$ in the \textit{clean}
case (elastic mean free path $l_{el} \gg v_{F}/\pi \Delta$) and
equals $\sqrt{8 D/\pi^{2} \Delta}$ in the \textit{dirty} case
($l_{el} \ll v_{F}/\pi \Delta$). Here $v_{F}$ denotes the Fermi
velocity and $D$ stands for the diffusion coefficient. The
function $f(x)$ is different for dirty and clean superconductors
but in the both cases it has the following asymptotic
\begin{gather}\label{f}
f(x) =
  \begin{cases}
    1, & \text{$x\gg 1$}, \\
    x, & \text{$x \ll 1$}.
  \end{cases}
\end{gather}

The result \eqref{1} holds as long as the superconductor and
ferromagnet are weakly coupled. The condition allows us to neglect
spin rotation by the exchange field induced in the superconductor
due to the proximity effect. The magnetization of a domain induces
the vector potential $A_{\rm eff}=H_{\rm ex}d$ where $d$ is the
characteristic size of a domain and $H_{\rm ex}$ is the exchange
field and, hence, the supercurrent at the superconductor near the
SF interface. The influence of the supercurrent on subgap electron
transport through the SF junction can be neglected if the
condition $e A_{\rm eff} d/hc\ll 1$ is hold that it is typically
so. Also we imply that the typical size of a domain is much larger
than the width of a domain wall, $d \gg \delta$. We leave more
complicated general case for future investigation.

\paragraph*{The model.} The hamiltonian describing a system of a
superconductor weakly coupled to a ferromagnet is as follows
\begin{gather}
H=H_S+H_F+H_{\rm int},
\end{gather}
where $H_S=\sum_{p,\sigma}E_{p\sigma}c_{p,\sigma}^\dag
c_{p,\sigma}+\sum_p \{\Delta c_{p,\uparrow}^\dag
c_{-p,\downarrow}^\dag+h.c.\}$ is the BCS hamiltonian of the
superconductor, $H_F=\sum_{k,\sigma} \varepsilon_{k,\sigma}
a_{k,\sigma}^{\dag} a_{k,\sigma}$ is the hamiltonian of the
ferromagnet and $H_{\rm
int}=\sum_{k,p,\sigma}\{a_{k,\sigma}^{\dag}
t_{k,\sigma;p,\sigma}c_{p,\sigma}+h.c.\}$. Here annihilation
operator $a_{k,\sigma}$ corresponds to the ferromagnet whereas
$c_{p,\sigma}$ to the superconductor. Labels $k$ and $p$ stand for
the momentums and symbol $\sigma=\pm 1$ denotes spin degree of
freedom.

The current flow through the SF junction can be described in terms
of the tunneling rates $\Gamma_{A}^{S\leftarrow F}(V)$ and
$\Gamma_{A}^{S\rightarrow F}(V)$. The first one has the meaning of
the probability per second for the Cooper pair creation in the
superconductor from two electrons with opposite spins in the
ferromagnet and vice-versa for $\Gamma_{A}^{S\rightarrow F}(V)$.
If the voltage between the superconductor and the ferromagnet is
less than the superconducting gap, $|eV|<\Delta$, the current
equals
\begin{gather}
\label{I} I(V)=e\left\{\Gamma_{A}^{S\leftarrow F
}(V)-\Gamma_{A}^{S\rightarrow F}(V)\right\}.
\end{gather}
By using the Fermi Golden rule the rates can be found in the
second order over the tunneling amplitude $t_{k,\sigma;p,\sigma}$.
Following the approach developed in Ref.~\cite{HekkingNazarov}
(and references therein) we finally obtain
%%%%%%%%%%%%%%%%%%%%%%%%%%%%%%%%%%%%%%%%%%%%%%%%%%%%%%%%%%%%%%%%%%%%%%%%%%%%%%%%%
%%%%%%%%%%%%%%%%%%%%%%%%%%%%%%%%%%%%%%%%%%%%%%%%%%%%%%%%%%%%%%%%%%%%%%%%%%%%%%%%%%%%5555
\begin{multline}
\label{Gamma} \Gamma_{A}^{S\leftarrow F}(V)=4\pi^3 \int d\xi
n_{F}(\xi-eV)n_{F}(-\xi-eV)\times
\\
\frac {\Delta^2}{[\Delta^2-\xi^2]}\sum_\sigma\tilde
\Xi_\sigma(2\sqrt{\Delta^2-\xi^2}),
\end{multline}
where $n_{F}(\xi)$ is the Fermi distribution function. Here and
below describing the calculations we believe that $\hbar=1$. The
$\Gamma_{A}^{S\rightarrow F}(V)$  can be obtained from the
expression for $\Gamma_{A}^{S\leftarrow F}(V)$ by substitution
$1-n_F$ for $n_F$. In derivation of Eq.\eqref{Gamma} we assumed
that the applied voltage is much smaller than the exchange energy,
$|eV|\ll E_{\rm ex}$, and the ferromagnet is strongly polarized,
$\nu_\parallel\gg\nu_\nparallel$. The conditions allow us to
neglect contributions to the conductance due to the interference
\cite{HekkingNazarov} in the ferromagnet.

The kernel $\tilde\Xi_{\sigma}(s)\equiv \int_0^\infty dt
\Xi_{\sigma}(t)e^{-st}$ is the Laplace transform of
$\Xi_{\sigma}(t)$. It can be expressed through the classical
probability $P(X_{1},\hat p_1; X_{2},\hat p_2,t)$ that an electron
with the momentum directed along $\hat p_1$ initially situated at
the point $X_1$ near the SF boundary arrives at time $t$ at some
point $X_2$ near the SF boundary with the momentum directed along
$\hat p_2$ spreading in the superconducting region as follows
\begin{multline}
\label{Xi} \Xi_\sigma(t)=\frac{1}{8 \pi^3 e^4 \nu_S}\int d\hat
p_{1,2}\int dX_{1,2} P(X_{1},\hat p_1; X_{2},\hat p_2,t)\times
\\
\left\{G(X_1,\hat p_1,\sigma)G(X_2,\hat p_2,
\sigma)\sin^2\left(\frac{\theta(X_1,X_2)} 2\right)+\right.
\\
+\left.G(X_1,\hat p_1,\sigma)G(X_2,\hat p_2,
-\sigma)\cos^2\left(\frac{\theta(X_1,X_2)} 2\right)\right\}.
\end{multline}
Here the spatial integration is performed over the surface of the
SF junction. We choose the spin quantization axis in the direction
of the local magnetization. The quasiclassical probabilities
$G(X,\hat p,\sigma)$ for the electron with spin polarization
$\sigma$ to tunnel from the ferromagnet to the superconductor are
normalized in such a way that the junction normal conductance per
unit area $g_{\sigma}(X)$ and the total normal conductance $G_{N}$
are determined as~\cite{HekkingNazarov,AverinNazarov}
\begin{gather}\label{g}
g_{\sigma}(X) = \int d\hat p G(X,\hat p,\sigma),\quad G_{N} = \int
d X \sum \limits_{\sigma} g_{\sigma}(X).
\end{gather}
Then the normal conductance per unit area discussed above is
defined as $g_N=G_N/{\cal A}$, where ${\cal A}$ is the surface
area of the SF interface. Symbol $\theta(X_1,X_2)$ is the angle
between the magnetizations of the ferromagnet at points $X_1$ and
$X_2$ near the SF boundary. In the limiting case of the monodomain
ferromagnet (so $\theta(X_1,X_2)\equiv 0$) and weak spin
polarization ($\nu_\parallel\approx\nu_\nparallel$)
Eqs.~\eqref{I}-\eqref{Xi} coincide with the results of Ref.
\cite{HekkingNazarov} and describe contribution to the subgap
conductivity of a superconductor -- normal metal junction due to
the interference in the superconductor. Eqs.\eqref{I}-\eqref{Xi}
describe the subgap current through the SF junction with general
domain structure of the ferromagnet. For the SF junction with the
monodomain fully polarized ferromagnet the subgap current vanishes
according to Eqs.\eqref{I}-\eqref{Xi}. However, inelastic
processes provide small but nonlinear contribution to the subgap
current which is asymmetric with the respect to the sign of the
bias voltage \cite{Falko}.

When the applied voltage is small, $|eV|\ll \Delta$, then the
current is proportional to the voltage, $I(V)=G_A V$. At low
temperature compared to the critical temperature of the
superconductor, $T\ll T_c$, Eqs.\eqref{I}-\eqref{Gamma} reduce to
\begin{gather}\label{GA}
G_A=8\pi^3e^2 \sum_\sigma \tilde\Xi_\sigma(2\Delta).
\end{gather}

\paragraph*{Andreev conductance of a single domain wall.}
The most interesting case is the fully polarized ferromagnet
because then the Andreev conductance of the SF junction is
completely determined by the contribution of domain walls. First
we consider the SF junction with the ferromagnet consisting of two
domains as shown in Fig.~\ref{fig1}. If we choose the frame of
reference according to Fig.~\ref{fig1} then the angle of
magnetization rotates as follows \cite{Landau8}
\begin{gather}\label{Bloch}
\theta(x,-\infty) = \arccos \tanh \frac{x}{\delta}, \\
\theta(x_{1},x_{2}) = \theta(x_{1},-\infty)-\theta(x_{2},-\infty).
\end{gather}
The classical probability $P(X_{1},\hat p_1; X_{2},\hat p_2,t)$ is
different in the dirty and clean superconductors. We consider
therefore these cases separately.

When the superconductor is \textit{dirty} we can neglect the
momentum dependence of the classical probability $P(X_{1},\hat
p_1; X_{2},\hat p_2,t)$, then
\begin{gather}
P(X_{1},X_{2}, t) =  \frac{2}{(4 \pi D t)^{3/2}} \exp \left [ -
\frac{(X_{1}-X_{2})^{2}}{4 D t} \right ], \label{PD}
\end{gather}
where the factor of $2$ appears because the superconductor
occupies the half-space. Now we can integrate over the momentum
directions $\hat p_{i}$ in Eq.(\ref{Xi}). Believing that
$g_{\sigma}(X)$ is a slow varying function of $X$ on the
lengthscale $\max\{\xi_{0},\delta\}$ we can perform the
integrations over the SF interface in Eq.(\ref{Xi}) and obtain
\begin{multline}\label{Dxi}
{\tilde \Xi}_{\sigma}(s) = \frac{L_{D}^{(\rm tot)} }{4 \pi^{4}
e^{4} \nu_{s} s} g_{\sigma} (g_{\sigma}-g_{-\sigma}) F_{d}\left (
\delta \sqrt
{\frac{s}{D}}\right ) +\\
+ \frac{{\cal A}}{8 \pi^{3} e^{4} \nu_{s} \sqrt{s D}} g_{\sigma}
g_{-\sigma} .
\end{multline}
Here the function $F_{d}(z)$ is defined as
\begin{gather}\label{F0}
F_{d}(z) = \int \limits_{0}^{\infty} dx K_{0}(x) x \tanh \left
(\frac{x}{2 z}\right ),
\end{gather}
where $K_{0}(x)$ is the modified Bessel function of the second
kind.

With a help of Eqs.~\eqref{GA} and \eqref{Dxi} we find that the
Andreev conductance of the SF junction can be written as
\begin{gather}\label{sum}
G_A=G_A^{(0)}+G_{A}^{(D)},
\end{gather}
where the surface and domain wall contributions are given by
\begin{gather}
G_A^{(0)}=\frac{4 {\cal A}}{ \pi e^{2} \nu_{s} \Delta \xi_{0}}
g_{\uparrow} g_{\downarrow},\label{GASD}
\\
G_{A}^{(D)}=\frac{L_{D}^{(\rm tot)} }{\pi e^{2} \nu_{s} \Delta}
 \left(g_{\uparrow}-g_{\downarrow}\right)^2 F_{d}\left ( \frac{4 \delta}{\pi \xi_{0}}\right ).
\label{GAWD}
\end{gather}
The surface contribution $G_A^{(0)}$ is suppressed in the case of
the fully polarized ferromagnet, $g_{\uparrow} \gg
g_{\downarrow}$, and the domain wall contribution $G_{A}^{(D)}$ is
the only that survives.

In the most interesting cases the function $F_{d}(z)$ has the
following asymptotic behavior
\begin{gather}\label{F1}
F_{d}(z) = \begin{cases} 1+ \displaystyle \frac{ \pi^{2}
z^{2}}{12} \left ( \ln z + 1 + \frac{6}{\pi^{2}}
\zeta^{'}(2)\right ), & \text{$z\ll 1$}, \\
    \displaystyle \frac{\pi}{4 z} \left ( 1 +
\frac{3}{4 z^{2}} \right ), & \text{$z\gg 1$}.
\end{cases}
\end{gather}
where $\zeta^{'}(z)$ is the derivative of the Riemann zeta
function. By using Eqs.\eqref{GA},\eqref{GAWD} and \eqref{F1} for
the case of the fully polarized ferromagnet, $g_{\downarrow}=0$,
we obtain the result \eqref{1}.

In the case of the {\it clean} superconductor we can estimate the
classical probability as
\begin{gather}
P(X_{1},X_{2}, t) =  \frac{2}{4 \pi (X_{1}-X_{2})^{2}} \delta
(|X_{1}-X_{2}|-v_{F} t). \label{PC}
\end{gather}
This probability describes tunneling through disordered SF
boundary. It allows to reproduce \cite{Chtch-Marenko_Nazarov} the
results of Ref.~ \cite{Falci_Feinberg_Hekking}.

In a similar way as above we obtain
\begin{multline}\label{Cxi}
{\tilde \Xi}_{\sigma}(s) = \frac{L_{D}^{(\rm tot)} }{4 \pi^{4}
e^{4} \nu_{s} s} g_{\sigma} ( g_{\sigma} - g_{-\sigma} ) F_{c}
\left (
\frac{2 \delta s}{v_{F}}\right ) +\\
+ \frac{\cal A }{8 \pi^{3} e^{4} \nu_{s} v_{F}} g_{\sigma}
g_{-\sigma} \left ( \ln \frac{v_{F}}{\lambda_{F} s} - \gamma
\right ),
\end{multline}
where $\gamma\approx 0.577$ is the Euler constant and the function
$F_{c}(z)$ is defined as
\begin{equation}\label{W}
F_{c}(z) = \int \limits_{0}^{\infty}d x K_{0}(x) \ln \cosh
\frac{x}{z}.
\end{equation}
Then, the surface and domain wall contributions to the Andreev
conductance are as follows
\begin{gather}
G_A^{(0)}=\frac{2 {\cal A}}{ \pi e^{2} \nu_{s}\Delta \xi_{0} }
g_{\uparrow} g_{\downarrow}\left ( \ln \frac{\xi_0}{\lambda_{F}} -
\gamma \right ), \label{GASC}
\\
G_{A}^{(D)}=\frac{L_{D}^{(\rm tot)} }{e^{2} \nu_{s} \Delta}
 \left(g_{\uparrow}-g_{\downarrow}\right)^2 F_{c}\left (
\frac{4 \delta}{\pi \xi_{0}}\right ),\label{GAWC}
\end{gather}
where $\lambda_{F}$ denotes the Fermi length. The function
$F_{c}(z)$ has the following asymptotic behavior
\begin{gather}\label{F2}
F_{c}(z) =\begin{cases}
     1-\displaystyle \frac{ \pi z}{ 2} \ln 2 , & \text{$z\ll 1$}, \\
    \displaystyle \frac{\pi}{4 z} \left ( 1 +
\frac{3}{z^{2}} \right ), & \text{$z\gg 1$},
\end{cases}
\end{gather}

For the case of the fully polarized ferromagnet,
$g_{\downarrow}=0$, Eqs.\eqref{GA},\eqref{GAWC} and \eqref{F2}
lead to the result \eqref{1}.

\paragraph*{Andreev conductance of several domain walls.}
Now we consider the domain structure with several domain walls
touching the SF interface. If domain walls separate domains with
the opposite directions of magnetization then the Andreev
conductance is a sum of contributions from each domain wall. By
assuming that the characteristic domain size is much larger than
the domain wall width and the magnetization rotation is given by
Eq.\eqref{Bloch}, we find the result \eqref{1} with $L_{D}^{(\rm
tot)}$ being the total length of the domain walls at the SF
interface. Usually, the domain structure at the SF interface is
more complicated. Nevertheless, the Andreev conductance remains
proportional to the total domain wall length whereas the function
$f(x)$ (see Eq.\eqref{1}) may depend on the particular domain
structure.

Possible experimental setup can be prepared in a similar way as in
recent experiment \cite{Ryazanov_vortices}. The normal conductance
between the superconductor and the ferromagnet should be smaller
than the normal conductances of the leads and the ferromagnet
(superconductor). The condition allows to neglect voltage
gradients in  the ferromagnet and believe that the ferromagnet is
in equilibrium and that we measure The Andreev conductance of the
SF junction rather than lead-related effects. By changing the
applied magnetic field we would change the number of domains in
the ferromagnet. As Eq.(1) shown, the Andreev conductance is
proportional to the number of domain walls in the ferromagnet and,
consequently, to the number of domains. It can be checked
experimentally by measuring the Andreev conductance as a function
of the applied magnetic field.
%We present the schematic dependence in Fig.~\ref{fig2}.

In conclusion, we evaluated the low-voltage Andreev conductance of
the SF junction when the ferromagnet is strongly polarized and
consists of several domains. The main transport mechanism under
subgap conditions is two electron tunneling (with zero total spin
of an electron pair) whereas the transfer of single electrons is
strongly suppressed.  Exchange field of the ferromagnet aligns
electron spins and suppresses the two electron tunneling. However
the tunneling is not suppressed near the domain walls where
electrons involved come from (or come to) the adjacent domains. It
is found that at strong polarization of the ferromagnet the domain
wall contribution to the Andreev conductance is the largest. We
presented an approach which gives an opportunity to find the
subgap current in wide range of layouts. Dynamics of domains with
magnetic field can be probed experimentally through the SF
conductance measurement.

We are especially grateful to V.V. Ryazanov for suggesting the
problem treated in this letter and stimulating discussions.
Furthermore we thank M.V. Feigelman, A.S. Iosselevich,  S. Koshuba
and Ya. Fominov for useful discussions. We wish to thank RFBR
(projects No. 03-02-06259, 03-02-16677, and 03-02-17494), the
Netherlands Organization for Scientific Research (NWO), the Swiss
NSF, Forschungszentrum J\"ulich (Landau Scholarship), the programs
of the Russian Ministry of Science: Mesoscopic systems and Quantum
computations and the program of the leading scientific schools
support.

%%%%%%%%%%%%%%%%%%%%%%%%%%%%%%%%%%%%%%%%%%%%%%%%%%%%%%%%%%%%%%%%%%%%%%%%%%%%%%%%%%%

\end{document}